\documentclass[aps,onecolumn,showpacs]{revtex4}
\usepackage{graphicx}
\usepackage{epsfig}
\usepackage{amssymb}
\usepackage{graphicx}
\usepackage{amsfonts}
\usepackage{latexsym}
\usepackage{amsmath}
\usepackage{mathrsfs}
\numberwithin{equation}{section}

\newcommand\encadremath[1]{\vbox{\hrule\hbox{\vrule\kern8pt
\vbox{\kern8pt \hbox{$\displaystyle #1$}\kern8pt}
\kern8pt\vrule}\hrule}}
\def\enca#1{\vbox{\hrule\hbox{
\vrule\kern8pt\vbox{\kern8pt \hbox{$\displaystyle #1$} \kern8pt}
\kern8pt\vrule}\hrule}}

\begin{document}
\title{Skew-orthogonal polynomials: the quartic case.}
\author{Saugata Ghosh}\email{sghosh@ictp.it}
\affiliation{F 253, Sushant Lok, Part II, Sector 57, Gurgaon,
India.}
\date{\today}
%\maketitle
\begin{abstract}
We present an iterative technique to obtain skew-orthogonal
polynomials with quartic weight, arising  in the study of symplectic
ensembles of random matrices.
\end{abstract}


\maketitle

\section{Introduction}

After the publication of \cite{ghosh1} and \cite{ghosh2}, late Prof.
M. L. Mehta was not particularly happy with the formal nature of the
presentation. Following his suggestion, this paper attempts to give
an explicit workout of the skew-orthogonal polynomials (SOP)
corresponding to the quartic potential.

We study SOP arising in the study of symplectic ensembles of random
matrices. The corresponding weight function is $w(x)=\exp[-V(x)]$,
where
\begin{equation}
\label{quartic}
 V(x)=\frac{x^4}{4}+\frac{\alpha x^2}{2},\qquad \alpha\in {\mathbb
 R},
\end{equation}
is the quartic potential. In this paper, we outline an iterative
technique to develop these polynomials which can be used to obtain
the level-density and $2$-point function for the symplectic
ensembles of random matrices. Here, we  emphasize that this method
can be easily extended to all potentials of the form
\begin{equation}
\label{vd} V(x)=\sum_{k=1}^{d}\frac{u_{2k}x^{2k}}{2k},\qquad
u_{2d}=1.
\end{equation}
Without loss of generality, we will be dealing with monic SOP of the
form
\begin{equation}
\label{phi} \phi_{n}(x)=w(x)\sum_{k=0}^{n}c_{k}^{(n)}x^{k},\qquad
c_{n}^{(n)}=1,
\end{equation}
and define
\begin{equation}
\label{phipsi} \psi_{n}(x):=\frac{d}{dx}\phi_{n}(x).
\end{equation}
They satisfy the skew-orthonormalization relation
\begin{equation}
\label{orthonorm}
 \int_{\mathbb
R}\phi_{n}(x)\psi_{m}(x)dx=g_{n}Z_{nm},
\end{equation}
where
\begin{eqnarray}
Z &=& \left(\begin{array}{cc}
0 & 1     \\
-1 & 0         \\
\end{array}\right)
\dotplus \ldots \dotplus,
\end{eqnarray}
is an anti-symmetric block-diagonal matrix with $Z^2=-1$. $g_{n}$ is
the normalization constant with the property
\begin{equation}
g_{2n}=g_{2n+1}.
\end{equation}
Here $\psi_{n}(x)$ is a polynomial of order $n+2d-1$ with leading
coefficient $-1$.  We have dropped the superscript $\beta$ (as in
\cite{ghosh1}) and \cite{ghosh2}) since we will only be interested
in SOP arising in the study of symplectic ensembles of random
matrices.

As explained in Ref.\cite{ghosh1} and \cite{ghosh2}, we expand
$\psi(x)$ in terms of $\phi(x)$ and write
\begin{equation}
x\psi_{n}(x)=\sum_{m=j}^{k} R_{nm}\phi_{m}(x),
\end{equation}
where from Eqs.(\ref{vd}, \ref{phi}, \ref{phipsi}), one can say that
$k=n+2d$. Furthermore, using the integral $\int_{\mathbb R}
x\psi_{n}(x)\psi_{m}(x)dx$, we can show that the matrix $R$ satisfy
anti-self dual relation

\begin{equation}
\label{dual}
 R=ZR^tZ\equiv-R^D.
\end{equation}
Summing up these results, we can write for any polynomial weight

\begin{eqnarray}
\label{drecursion}
 x\psi_{2n}(x) &=&
R_{2n,2n+2d}\phi_{2n+2d}(x)+\ldots +
R_{2n,2n-2d}\phi_{2n-2d}(x),\\
\label{drecursion1} x\psi_{2n+1}(x) &=&
R_{2n+1,2n+2d+1}\phi_{2n+2d+1}(x)+\ldots +
R_{2n+1,2n-2d}\phi_{2n-2d}(x),
\end{eqnarray}
where $\phi_{-m}=0$, $m$ being a positive integer. Our choice of the
potential ($u_{2d}=1$) ensures that
\begin{equation}
\label{highest}
 R_{m,m+2d}=-1,\qquad m\geq 0.
\end{equation}
 Also from Eq.(\ref{dual}), we can show
that
\begin{eqnarray}
\label{lowest}
\nonumber
 R_{2n,2n-2d} = - R_{2n-2d+1,2n+1} &=& 1,\qquad n\geq d,\\
\nonumber
                                   &=& 0,\qquad n<d,\\
\nonumber
R_{2n+1,2n-2d+1} = - R_{2n-2d,2n}  &=& 1,\qquad n\geq d,\\
                                   &=& 0, \qquad n<d.
\end{eqnarray}

Following this brief recapitulation of \cite{ghosh1} and
\cite{ghosh2}, we will outline our plan of action.

1. We will calculate the SOP's $\phi_{n}(x)$ for $0\leq n <2d$ (for
the particular case $d=2$) using generalized Gram-Schmidt
orthogonalization.

2. Using these polynomials, we will use Eqs.(\ref{drecursion}) and
(\ref{drecursion1}) to calculate the higher order polynomials
recursively. The coefficients will be expressed in terms of
integrals of the form $\int_{\mathbb R} x^n\exp[-2V(x)]dx$.

3. We show that these coefficients themselves satisfy a set of
difference equations.

4. Using these difference equations, we also obtain the
normalization constants.

5. Finally, we talk briefly about the zeros of these
polynomials.

\section{The Gram-Schmidt technique}

Putting $n=0$ in  Eqs.(\ref{drecursion}) and (\ref{drecursion1})
(keeping in mind Eq.(\ref{lowest})), we can see that it is useless,
unless we have information about the SOP's  $\phi_{m}$ for $2d>m\geq
0$. To overcome this problem, we will obtain these $2d$ polynomials
$(\phi_{0}(x),\ldots,\phi_{2d-1}(x))$ using the Gram-Schmidt
technique for SOP. From here onward, we will focus our attention on
the specific weight function defined in Eq.(\ref{quartic}), although
the method outlined can be extended to any $d$. For $d=2$, the first
$4$ monic polynomials can be written as

\begin{eqnarray}
\phi_{0}(x)&=& w(x) , \qquad \phi_{1}(x) = xw(x),\\
\phi_{2}(x) &=& (x^2+c_{0}^{(2)})w(x),\qquad \phi_{3}(x) =
(x^3+c_{1}^{(3)}x)w(x).
\end{eqnarray}
Correspondingly
\begin{eqnarray}
\psi_{0}(x)&=& -V'(x)w(x) , \qquad \psi_{1}(x) = (1-xV'(x))w(x),\\
\psi_{2}(x) &=& [2x-(x^2+c_{0}^{(2)})V'(x)]w(x),\qquad \psi_{3}(x) =
[3x^2+c_{1}^{(3)}-V'(x)(x^3+c_{1}^{(3)}x)]w(x).
\end{eqnarray}
Using
\begin{eqnarray}
\int_{-\infty}^{\infty}\phi_{2}(x)\psi_{1}(x)dx=0,
\end{eqnarray}
we get
\begin{eqnarray}
C_{0}^{(2)}=-\frac{\int_{-\infty}^{\infty}
x^2(1-xV'(x)){w^{2}(x)}dx}{\int_{-\infty}^{\infty}{(1-xV'(x))w^{2}(x)}dx}.
\end{eqnarray}
Similarly, using
\begin{eqnarray}
\int_{-\infty}^{\infty} \phi_{3}(x)\psi_{0}(x)dx=0,
\end{eqnarray}
we get
\begin{eqnarray}
C_{1}^{(3)}=-\frac{\int_{-\infty}^{\infty} x^3
V'(x){w^{2}(x)}dx}{\int_{-\infty}^{\infty} xV'(x){w^{2}(x)}dx}.
\end{eqnarray}
We also get
\begin{eqnarray}
\label{g0}
 g_{0}=g_{1}=\int_{-\infty}^{\infty}
\phi_{0}(x)\psi_{1}(x)dx,\qquad
g_{2}=g_{3}=\int_{-\infty}^{\infty}\phi_{2}(x)\psi_{3}(x)dx.
\end{eqnarray}

\section{Recursion relation}
Now, let us look at Eqs.(\ref{drecursion}) and (\ref{drecursion1}).
Since $V(x)=V(-x)$, we have
\begin{eqnarray}
\psi_{2n}(x)=-\psi_{2n}(-x),\qquad \psi_{2n+1}(x)=\psi_{2n+1}(-x).
\end{eqnarray}
Thus the odd (even) terms will  be absent in Eqs.(\ref{drecursion})
((\ref{drecursion1})) since
\begin{eqnarray}
\nonumber   R_{2n,2n+2k+1} =
-\frac{1}{g_{2n+2k}}\int_{-\infty}^{\infty}x\psi_{2n}(x)\psi_{2n+2k}(x)dx
= 0,
\end{eqnarray}
and
\begin{eqnarray}
 R_{2n+1,2n+2k} =
\frac{1}{g_{2n+2k}}\int_{-\infty}^{\infty}x\psi_{2n+1}(x)\psi_{2n+2k+1}(x)dx
= 0.
\end{eqnarray}
Using these results and Eq.(\ref{highest}), we can rewrite Eqs.
(\ref{drecursion}) and (\ref{drecursion1}) for $n=0$ and $d=2$ as

\begin{eqnarray}
\label{4recursion1}
x\psi_{0}(x) &=& -\phi_{4}(x)+R_{0,2}\phi_{2}(x)+R_{0,0}\phi_{0}(x),\\
\label{4recursion2}
 x\psi_{1}(x) &=&
-\phi_{5}(x)+R_{1,3}\phi_{3}(x)+R_{1,1}\phi_{1}(x).
\end{eqnarray}
Also, using the skew-orthogonality property, we get
\begin{eqnarray}
\label{r00}
R_{0,0} &=& \frac{1}{g_0}\int_{-\infty}^{\infty} x\psi_{0}(x)\psi_{1}(x)dx, \\
\label{r02}
 R_{0,2} &=& \frac{1}{g_2}\int_{-\infty}^{\infty} x\psi_{0}(x)\psi_{3}(x)dx.
\end{eqnarray}
Similarly,
\begin{eqnarray}
\label{r11}
R_{1,1} &=& -\frac{1}{g_0}\int_{-\infty}^{\infty}
 x\psi_{0}(x)\psi_{1}(x)dx=-R_{0,0}, \\
\label{r13}
 R_{1,3} &=& -\frac{1}{g_2}\int_{-\infty}^{\infty} x\psi_{1}(x)\psi_{2}(x)dx.
\end{eqnarray}
Here, we note that $R_{0,0}=-R_{1,1}$ can be obtained directly from
Eq.(\ref{dual}). Once the coefficients are known, one can calculate
$\phi_{4}(x)$ and $\phi_{5}(x)$ using Eqs. (\ref{4recursion1}) and
(\ref{4recursion2}). Using (\ref{phipsi}), we can calculate
$\psi_{4}(x)$ and $\psi_{5}(x)$. We can also calculate
\begin{eqnarray}
g_{4}=g_{5}=\int_{-\infty}^{\infty} \phi_{4}(x)\psi_{5}(x)dx.
\end{eqnarray}
We will now calculate $\phi_{6}(x)$ and $\phi_{7}(x)$ using
\begin{eqnarray}
\label{4recursion11}
x\psi_{2}(x) &=& -\phi_{6}(x)+R_{2,4}\phi_{4}(x)+R_{2,2}\phi_{2}(x)+R_{2,0}\phi_{0}(x),\\
\label{4recursion22}
 x\psi_{3}(x) &=&
-\phi_{7}(x)+R_{3,5}\phi_{5}(x)+R_{3,3}\phi_{3}(x)+R_{3,1}\phi_{1}(x).
\end{eqnarray}
Again, using Eq.(\ref{dual}), we have
\begin{eqnarray}
R_{2,0}=-R_{1,3},\qquad R_{3,1}=-R_{0,2},\qquad R_{2,2}=-R_{3,3}.
\end{eqnarray}
$R_{0,2}$ and $R_{1,3}$ has already been calculated in (\ref{r02})
and (\ref{r13}) respectively. Also
\begin{eqnarray}
\label{r24}
R_{2,4} &=& \frac{1}{g_4}\int_{-\infty}^{\infty} x\psi_{2}(x)\psi_{5}(x)dx, \\
\label{r35}
 R_{3,5} &=& -\frac{1}{g_4}\int_{-\infty}^{\infty} x\psi_{3}(x)\psi_{4}(x)dx,
\end{eqnarray}
and
\begin{eqnarray}
\label{r22} R_{2,2}=-R_{3,3} &=&
\frac{1}{g_2}\int_{-\infty}^{\infty} x\psi_{2}(x)\psi_{3}(x)dx.
\end{eqnarray}
With these, we can calculate $\phi_{6}(x)$, $\phi_{7}(x)$ and
correspondingly $\psi_{6}(x)$ and $\psi_{7}(x)$ and have
\begin{eqnarray}
g_6=g_7=\int_{-\infty}^{\infty}\phi_{6}(x)\psi_{7}(x)dx.
\end{eqnarray}
Following the same technique, we can obtain the polynomials for all
$n$. For  $n\geq 2$ and $d=2$, we may rewrite Eqs.(\ref{drecursion})
and (\ref{drecursion1}) as

\begin{eqnarray}
\label{recursion4}
x\psi_{2n}(x)=\sum_{m=-2}^{2}R_{2n,2n+2m}\phi_{2n+2m}(x),\qquad
x\psi_{2n+1}(x)=\sum_{m=-2}^{2}R_{2n+1,2n+2m+1}\phi_{2n+2m+1}(x).
\end{eqnarray}
From Eq.(\ref{highest}), we have
\begin{equation}
\label{highest4}R_{2n,2n+4}=R_{2n+1,2n+5}=-1.
\end{equation}
Also from Eq.(\ref{lowest}), we have

\begin{eqnarray}
\label{lowest4}
 R_{2n,2n-4}=-R_{2n-3,2n+1}=1,\qquad
R_{2n+1,2n-3}=-R_{2n-4,2n}=1.
\end{eqnarray}
$R_{2n,2n-2}$ (for the even case) and $R_{2n+1,2n-1}$ (for the odd
case) is already known since
\begin{eqnarray}
\label{middle4}
 R_{2n,2n-2}=-R_{2n-1,2n+1},\qquad
R_{2n+1,2n-1}=-R_{2n-2,2n}.
\end{eqnarray}
Finally, we are left with terms of the form $R_{k,k+2}$ and
$R_{k,k}$ (for both $k$ odd and even). They are given by
\begin{eqnarray}
\label{eqr2n2k}
 R_{2n,2n+2k}=\frac{1}{g_{2k+2n}}\int_{-\infty}^{\infty}
x\psi_{2n}(x)\psi_{2n+2k+1}(x)dx,\qquad k=0,1\\
\label{eqr2n12k1}
 R_{2n+1,2n+2k+1}=-\frac{1}{g_{2k+2n}}\int_{-\infty}^{\infty}
  x\psi_{2n+1}(x)\psi_{2n+2k}(x)dx,\qquad k=0,1.
\end{eqnarray}
Here, we note that by expanding $x\psi_{2n}$ (or $x\psi_{2n+1}$), we
can evaluate $R_{2n,2n+2k}$ (or $R_{2n+1,2n+2k+1}$) analytically.
But that will involve terms like $c^{(2n+4)}_{2n+2}$ (or
$c^{(2n+5)}_{2n+3}$) and thereby cannot be used to obtain the
polynomials recursively. We might recall that for monic orthogonal
polynomials, this problem does not exist \cite{szego}. However it is
possible to evaluate Eqs.(\ref{eqr2n2k}) and (\ref{eqr2n12k1})
numerically, although it may be a tiresome process.

\section{The normalization constant}

So far, we have outlined a formalism to obtain the polynomials
recursively, where the recursion coefficients are expressed in terms
of certain integrals which needs to be evaluated at every iteration.
It is not practical to use this process for the study of large $n$
behavior of these polynomials. Nor is it convenient to study
potentials with larger $d$, since the number of terms in the
recursion relation increases with $d$.

In this section, we present an alternative technique to obtain both
the recursion coefficients and the normalization constant. We begin
with the identity (obtained by integration by parts)
\begin{eqnarray}
\label{identity}
 \int{\left[x(x\psi_{j}(x))'\right]}'\phi_{k}(x)dx =
-\int (x\psi_{k}(x)){(x\psi_{j}(x))}'dx,\qquad j,k=0,1,\ldots.
\end{eqnarray}

The only criteria to use this formalism is that as initial
condition, we need to know these polynomials for $n=0,\ldots,2d-1$.
Taking $j=2n$ and $k=2n+1$ in Eq.(\ref{identity}), we get the
recursion relation for the normalization constant.  We get

\begin{eqnarray}
\label{g4}
g_{4}+\gamma_{0}g_{2}-(1+\gamma_{0})g_{0} &=& 0,\qquad n=0,\\
\label{g6}
g_{6}+\gamma_{1}g_{4}-(1+\gamma_{0}+\gamma_{1})g_{2}+\gamma_{0}g_{0} &=& 0,\qquad n=1,\\
\label{g2n4}
g_{2n+4}+\gamma_{n}g_{2n+2}-(2+\gamma_{n}+\gamma_{n-1})g_{2n}+\gamma_{n-1}g_{2n-2}
+g_{2n-4}&=& 0,\qquad n\geq 2,
\end{eqnarray}
where
\begin{eqnarray}
\label{gamman}
 \gamma_{n}:=R_{2n,2n+2}R_{2n+1,2n+3}.
\end{eqnarray}

Here, we have expanded $x\psi_{2n}(x)$ and $x\psi_{2n+1}(x)$ (using
Eq.(\ref{recursion4})), and using Eqs.(\ref{highest4}),
({\ref{lowest4}}) and (\ref{middle4})), get the result. We have
assumed $\phi_{-n}$ and consequently $g_{-n}$ and $\gamma_{-n}$ is
zero, $n$ being a positive integer. Here, we must also remember that
for these SOP, $g_{2n}=g_{2n+1}$.

To evaluate $\gamma$, we need to know $R_{k,k+2}$ for both $k$ odd
and even. This can be calculated  from Eq.(\ref{identity}) by
putting
  $j=2n$, $k=2n+3$ and $j=2n+1$, $k=2n+2$. We get
\begin{eqnarray}
\label{rn35}
R_{2n+3,2n+5}(g_{2n+4}-g_{2n+2})-R_{2n+1,2n+1}R_{2n,2n+2}(g_{2n}-g_{2n+2})
+R_{2n-1,2n+1}(g_{2n-2}-g_{2n+2}) &=& 0,\\
\label{rn24}
R_{2n+2,2n+4}(g_{2n+4}-g_{2n+2})-R_{2n,2n}R_{2n+1,2n+3}(g_{2n}-g_{2n+2})
+R_{2n-2,2n}(g_{2n-2}-g_{2n+2}) &=& 0.
\end{eqnarray}

These three equations can be used together to obtain $g_k$ for all
$k$. For example, a knowledge of $R_{0,2}$ and $R_{1,3}$ (which we
get from Eqs.(\ref{r02}) and (\ref{r13})) will give $\gamma_0$. This
will give $g_{4}$ from (\ref{g4}) (again we need $g_0$ and $g_2$
which we will extract from Eq.(\ref{g0})). A knowledge of $g_4$ in
turn yields $R_{3,5}$ and $R_{2,4}$ (where we use  $R_{0,0}$ and
$R_{1,1}$, calculated from Eq.(\ref{r11})) from Eqs.(\ref{rn35}) and
(\ref{rn24}), which gives $\gamma_1$. This in turn can be used to
calculate $g_6$ (\ref{g6}) and so on.

However, we still need to know $R_{j,j}$ for $j\geq 2$. This can be
obtained by writing $j=2n+3$ and $k=2n$ in Eq.(\ref{identity}). We
get

\begin{eqnarray}
\label{rkk}
R_{2n+3,2n+3}R_{2n,2n+2}(g_{2n+2}-g_{2n})+R_{2n+3,2n+5}(g_{2n}-g_{2n+4})
+R_{2n-1,2n+1}(g_{2n}-g_{2n-2})=0.
\end{eqnarray}
$R_{2n+2,2n+2}$ can be  calculated from the relation
$R_{2n+2,2n+2}=-R_{2n+3,2n+3}$. Thus a knowledge of $g_4$, $R_{0,0}$
and $R_{0,2}$ gives $R_{3,5}$ (\ref{rn35}). Then using $g_4$, $g_2$,
$R_{3,5}$ and $R_{0,2}$, we can obtain $R_{3,3}$ (and hence
$R_{2,2}$) from (\ref{rkk}).

To summarize, we have outlined a recursive technique to obtain the
normalization constant. This necessitates a knowledge of
coefficients of the form $R_{k,k+2}$. Again to evaluate this (also
needed in Eq.(\ref{recursion4})), we need $R_{k,k}$. We obtain a
series of self-consistent recursion relations to obtain these
coefficients.

{\bf Comment:} In the context of random matrix theory, we may point
out that now we can obtain various statistical properties like the
level-density and $2$-point functions for even finite dimensional
symplectic ensembles of random matrices (\cite{ghosh1} \cite{ghosh3}
\cite{puri}).

\section{The Zeros of these polynomials}
Having obtained the  polynomials for the specific case of $d=2$, we
will discuss briefly about the zeros of these SOP in this section.
We know that in a given interval $[a,b]$

\begin{eqnarray}
\forall n\neq 0, \qquad \int_{a}^{b}\psi_{n}(x)w(x)dx=0,\qquad n\geq
2.
\end{eqnarray}
So $\psi_{n}(x)$ should have atleast one point in the interior of
$[a,b]$ where it changes sign. Let there be $k$ such points
$x_1,x_2,\ldots,x_k$. Then the function $\psi_{n}(x)(x-x_1)\ldots
(x-x_k)$ is positive or negetive definite for $k\leq n+2d-1$, since
$\psi_{n}(x)$ is a polynomial of order $n+2d-1$. This implies
\begin{eqnarray}
\int_{a}^{b}\psi_{n}(x)(x-x_1)\ldots (x-x_k)dx \neq 0, \qquad
\forall k\leq n+2d-1.
\end{eqnarray}
However, from skew-normalization relation, this condition is
satisfied if and only if

1. For $n=2m$, $k=2m+1$.

2. For $n=2m+1$, $k=2m$.

3. $(x-x_1)\ldots (x-x_k)=\phi_{k}(x)$.

This implies that

a. $\phi_{k}(x)$ is a polynomial of order $k$ with $k$ real zeros.

b. $\psi_{k}(x)$ is a polynomial of order $k+2d-1$, with
$\psi_{2m}(x)$ having $2m+1$ and $\psi_{2m+1}(x)$ having $2m$ real
zeros.

\section{Conclusion}
This paper gives a formalism to derive the SOP arising in the study
of symplectic ensembles of random matrices \cite{mehta}. Here, we
would like to point out that this paper contains almost no new
results which were not present in \cite{ghosh1} and \cite{ghosh2}.
However, a lot of the properties were overlooked. For example, the
explicit form of the recursion coefficients, the necessity to use
Gram-Schmidt method to obtain polynomials of order $n<2d$, the zeros
of $\phi_{n}(x)$ and $\psi_{n}(x)$ etc. At this point, one can
easily obtain these polynomials, although a deeper insight into the
recursion coefficients (specially its large $n$ behavior) is an
absolute necessity. However, with all its limitations, we hope that
this paper will take us further towards our goal, which is ``to
develop the theory of skew orthogonal polynomials until it becomes a
working tool as handy as the existing theory of orthogonal
polynomials'' \cite{dyson}.

{\bf Acknowledgment}: The author is grateful to Mr. T. K. Bhaumik
for his support during the completion of this work.


\begin{thebibliography}{1}

\bibitem{ghosh1}Ghosh S., 2006,
Generalized Christoffel-Darboux formula for skew-orthogonal
polynomials and random matrix theory,
 J. Phys. A: Math. Gen. {\bf 39} 8775-8782.\\

\bibitem{ghosh2}Ghosh S., 2007,
 Skew-orthogonal polynomials, differential systems and random
matrix theory, J. Phys. A : Theor. {\bf 40}.


\bibitem{ghosh3} Ghosh S., 2004,
 Long-range interactions in the quantum many-body problem in one
dimension: Ground state, Phys. Rev. E 69, 036118 (2004)

\bibitem{puri}Ghosh S., Pandey A, Puri S., and Saha R, 2003,
Non-Gaussian random-matrix ensembles with banded spectra, Phys. Rev.
E 67, 025201 (2003)

\bibitem{szego} {\it Orthogonal Polynomials}, Szego G., 1939
(American Mathematical Society, Providence)

\bibitem{mehta}
{\it Random Matrices}, Mehta M. L., 2004 (The Netherlands, Elsevier,
3rd ed.)


\bibitem{dyson} Dyson F. J., 1972, A Class of Matrix Ensembles, J.
Math. Phys.  {\bf 13}, 90-97.


\end{thebibliography}
\end{document}